\documentclass{ledger}
\include{epsfig}

\newcommand{\thefirstpagenum}[0]{1}
\newcommand{\thelastpagenum}[0]{5}
\newcommand{\theyear}[0]{201X}
\newcommand{\thevol}[0]{X}
\newcommand{\thedoi}[0]{DOI 10.5195/LEDGER.\theyear.X}
\newcommand{\ledgerpages}[0]{\thefirstpagenum-\thelastpagenum}

\hypersetup{pdfauthor={Reginald Smith}, pdftitle={Bitcoin Average Dormancy}}

\title{Bitcoin Average Dormancy: A Measure of Turnover and Trading Activity}
\author{Reginald D. Smith\thanks{R.~D.~Smith (rsmith@supremevinegar.com) is an independent researcher in the fields of the physics of complex systems, quantitative linguistics, and supply chains. He studied physics at the University of Virginia and received a MBA from the MIT Sloan School of Management. He has studied and owned Bitcoin since 2014. He is currently owner and president of Supreme Vinegar LLC, a vinegar manufacturer in Pennsylvania. \newline 1PkoFCwKikHQgFtsHtfNvkz8bZ8w7Hsmdr}}

\pretitle{
  \centering \selectfont LEDGER \LaTeX \ TEMPLATE \par
  \centering \selectfont RESEARCH ARTICLE \par
  \fontsize{24pt}{28pt}\selectfont} 

\begin{document}

\maketitle

\thispagestyle{pagefirst}

\begin{abstract}
Attempts to accurately measure the monetary velocity or related properties of bitcoin used in transactions have often attempted to either directly apply definitions from traditional macroeconomic theory or to use specialized metrics relative to the properties of the Blockchain like bitcoin days destroyed. In this paper, it is demonstrated that beyond being a useful metric, bitcoin days destroyed has mathematical properties that allow you to calculate the average dormancy (time since last use in a transaction) of the bitcoins used in transactions over a given time period. In addition, bitcoin days destroyed is shown to have another unexpected significance as the average size of the pool of traded bitcoins by virtue of the expression Little's Law, though only under limited conditions.
\end{abstract}

\section{Introduction}

Since the now legendary white paper by Satoshi Nakamoto \cite{satoshi} in 2008 and Bitcoin's later launch in January 2009, Bitcoin has rapidly emerged to become one of the most radical and ingenious financial innovations in recent times. Bitcoin's success relied on solving several previous problems with past digital currencies including centralization, trust, and transaction ledgers. Bitcoin is based on a decentralized network of all past transactions (the Blockchain) which is assembled from units of multiple transactions called blocks. Miners bundle transactions into blocks, recognized by a network consensus, upon solving a Proof of Work (hashing computation), in exchange for newly mined Bitcoin. Hundreds of other cryptocurrencies, termed alt-coins, have also evolved using most of the same basic concepts. This paper assumes a working knowledge of Bitcoin and good introductions of its basic concepts are given in \cite{bitcoin1,bitcoin2,bitcoin3}. In addition, interesting comments on Bitcoin's various descriptions as both a currency and a store of value can be found in \cite{store,store2}.

As Bitcoin grew and evolved after 2009, it became increasingly important to have metrics to characterize the growth and behavior of Bitcoin based off of data easily obtainable from analysis of the Blockchain. Amongst the first, and most obvious, was the measurement of total transaction volume in bitcoin (BTC) on-chain (transactions recorded in the Blockchain). This measure had many drawbacks, however, one of those being that it did not differentiate between transactions that may be trivial, such as users moving bitcoin between several addresses they own, and more substantive transactions reflecting the acceptance of bitcoin and usage in the wider economy.

In order to deal with this issue, the idea of “bitcoin days destroyed” was first proposed on the forum Bitcointalk.org on April 20, 2011 by the user ByteCoin \cite{ddorigin}. Bitcoin days destroyed would be a weighted measure of transaction volume where transactions in BTC would be multiplied by the number of days since those bitcoin were last spent.

The advantage of bitcoin days destroyed was that the data to calculate it was easily accessible in the input data for every transaction. Bitcoin days destroyed would give a more reliable measure of economic activity in Bitcoin since it would help indicate when long dormant coins were used again for transactions and weighted accordingly. So 1 BTC last spent 100 days prior produces an equivalent days destroyed to  100 BTC spent one day after their previous transaction. The heavier weighting for less frequently circulating coins helps remove some `noise' that is due to rapid, repeating short term transactions that may not be indicative of the true demand for Bitcoin usage. It has been viewed by many as both a better indicator of relative economic activity than raw transaction volume and also a proxy for monetary velocity \cite{velocity1, velocity2, velocity3}. While several studies have referenced bitcoin days destroyed as a measure of economic activity or liquidity \cite{ddstudy1,ddstudy2}, no studies have yet analyzed bitcoin days destroyed itself in detail.

The analysis of bitcoin days destroyed in a fuller context is important since it is not just a good metric, it has integral links to the macroscale usage of bitcoin and is a deep insight into the development of the bitcoin economy and user behavior. Therefore, placing bitcoin days destroyed in a mathematical and more general framework will help us not just understand bitcoin days destroyed better but understand the past and evolution of Bitcoin in a new light.

\subsection{Days destroyed and its relation to bitcoin transaction volume}

For a given bitcoin transaction, bitcoin days destroyed is defined as the value of the transaction in BTC times the integer number of days since these bitcoins were previously sent in a transaction. Note that in a transaction, not all bitcoin spent were necessarily last spent the same amount of time ago and for the purpose of calculating bitcoin days destroyed all BTC are weighted separately by their last spent date as indicated in the blockchain. For a set of bitcoin transactions, for example all transactions within a 24 hour period, the total bitcoin days destroyed can be defined as the sum of bitcoin days destroyed across all transactions in the set.

Let us define as $\mathcal{S}$ as the set of all transactions within a given period of time. For the $i$th transaction in $\mathcal{S}$, let the value of the transaction in BTC be represented as $b_i$ and the number of days since these bitcoin were previously transacted as $\Delta t_i$. The total value of bitcoin days destroyed for all transactions, $D$, is given by

\begin{equation}
D=\sum_{i=1}^{\mathcal{S}} d_i = \sum_{i=1}^{\mathcal{S}} b_i \Delta t_i
\label{ddeq1}
\end{equation}

If the total value of all bitcoin transacted in $\mathcal{S}$ is designated by $B$, we can further state

\begin{equation}
D=B\sum_{i=1}^{\mathcal{S}} \frac{b_i}{B} \Delta t_i
\label{ddeq2}
\end{equation}

The fraction $\frac{b_i}{B}$ is the weighted proportion of total transactions represented by the $i$th transaction. Therefore, the sum in equation \ref{ddeq2} is essentially equal to the weighted average number of days since the last transaction across all transactions. We will designate this average time as the average dormancy, $\langle t \rangle$ which is related to both the transaction volume and bitcoin days destroyed

\begin{equation}
D= B \langle t \rangle
\label{ddeq3}
\end{equation}

The average dormancy, easily calculated from bitcoin transaction volume and bitcoin days destroyed, is interesting in several respects. First, and most obvious, it links the two most important measures of Bitcoin economic activity together in a more straightforward manner than the algorithm to calculate days destroyed based off the transaction volume. Second, it gives us a rough idea on how long coins involved in current transactions are being left unused between transactions.

To understand average dormancy, a key example may be helpful. Take a Bitcoin user, Jill who receives 10 BTC on day 0. On day 5, she decides to make transactions sending 2 BTC each to Anya, Bob, Cai, Dave, and Elena. A total of 10 BTC created 5 days ago is spent creating 50 bitcoin days destroyed. Next, each of the five recipients spends 2 BTC staggered by one day per person. Thus we have 2 x (1+2+3+4+5) = 30 bitcoin days destroyed. From the time of Jill's first receipt of her bitcoins, on average, how long were these bitcoins dormant between spending periods? The answer is now easy. Over the entire period, 80 bitcoin days were destroyed. During the same time, 20 BTC in total transactions occurred. Thus $\langle t \rangle =$ 80 bitcoin days destroyed / 20 BTC transaction volume = 4 days. This is in agreement with the fact the first time 10 BTC was spent, the bitcoins were dormant for 5 days while the second period over which 10 BTC was spent, the bitcoins were dormant an average of 3 days.

While the average dormancy is not a true monetary velocity measure since it does not take the entire money supply or price levels into account as in the exchange equation, it can give us a good idea about the movement and circulation of those bitcoins that are actively being used in economic activity. Valuably, this metric correlates with other indicators, particularly the exchange rate of Bitcoin with fiat (USD/BTC) and allows us to test hypotheses on the behavior of bitcoin users, especially hoarders, under various economic conditions. In addition, the inverse of the average dormancy multiplied by a period of time (say 90 days or 365 days), which we will term turnover, tells us the average number of times the actively used bitcoins can be expected to be spent in on-chain transactions during that time, given $\langle t \rangle$ is the time to be spent once on average.

\section{Average dormancy over time}

The sources analyzed in this paper all came from the historical variables from Blockchain.info \cite{blockchain}, obtained indirectly through Quandl. The exception is all measures of days destroyed which came from the bitcoin transaction data portal OXT \cite{oxt}.  The Blockchain.info measure of transaction volume in BTC used is the Estimated Transaction Value that subtracts transactions of change returned to senders in transactions.

\begin{figure}[ht]
\centering
 \begin{tabular}{cc}
 	 \includegraphics[height=2.5in, width=2.5in]{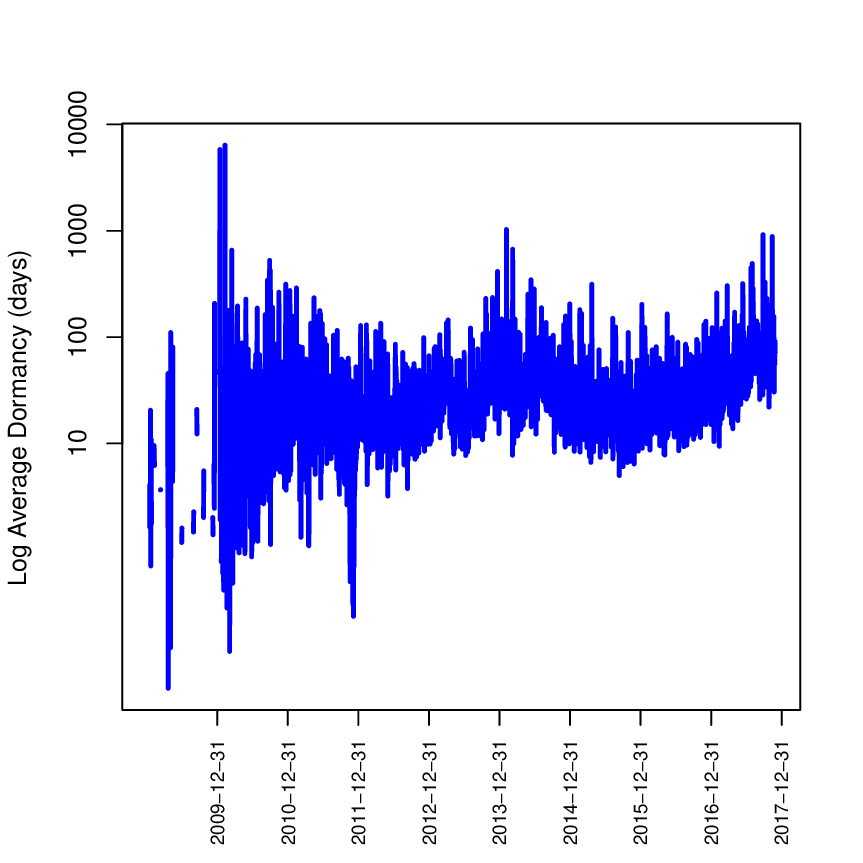}&
    \includegraphics[height=2.5in, width=2.5in]{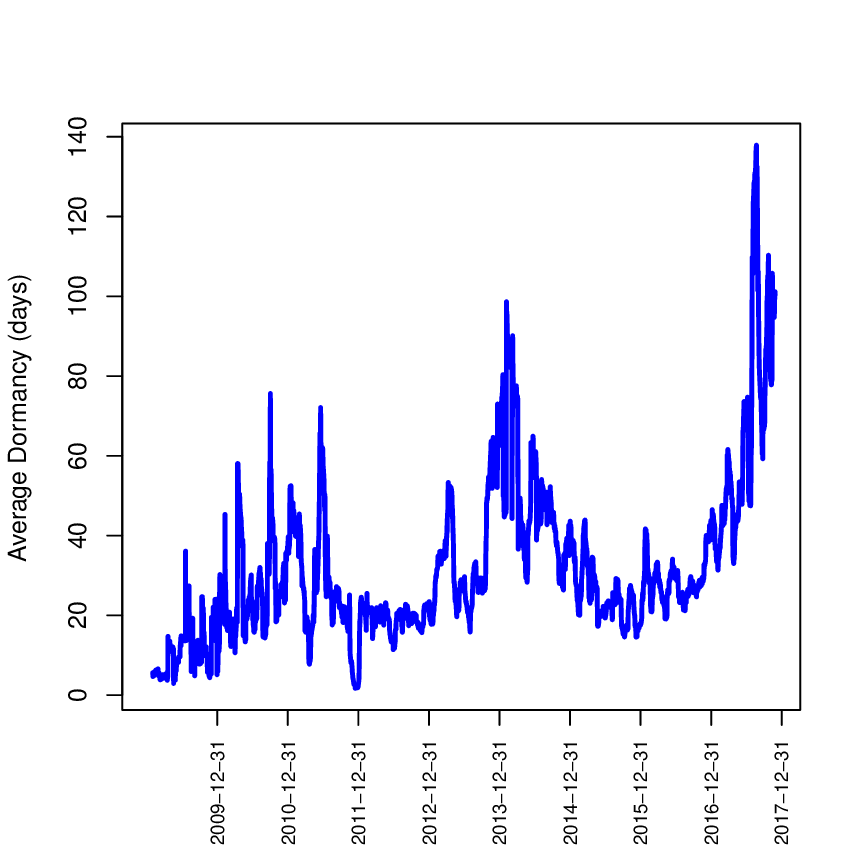}\\
\includegraphics[height=2.5in, width=2.5in]{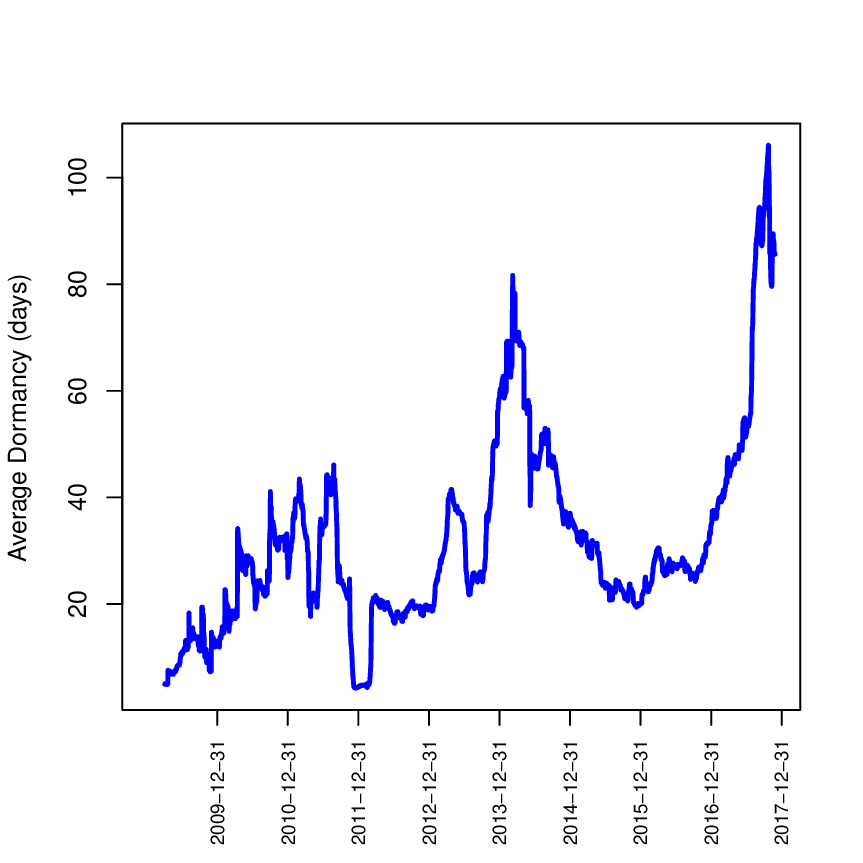}\\
 \end{tabular}
\caption{Average dormancy over time using daily data, 30 day aggregated transactions, and 90 day aggregated transactions.}
\label{dormfig}
\end{figure}

\begin{figure}[ht]
\centering
 \begin{tabular}{cc}
 	 \includegraphics[height=2.5in, width=2.5in]{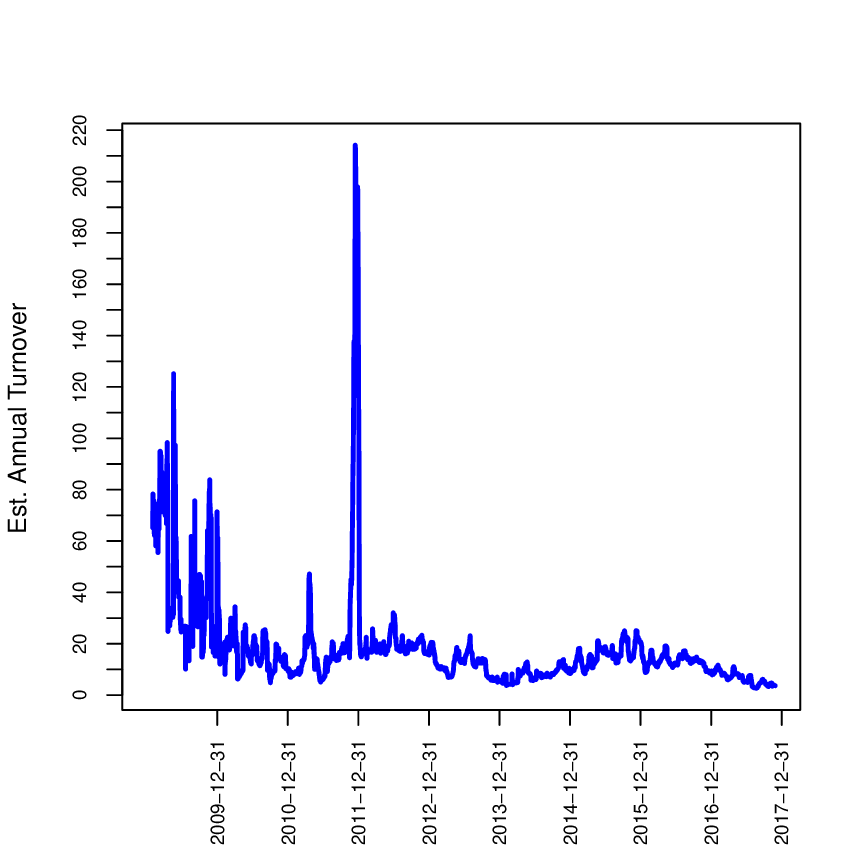}&
    \includegraphics[height=2.5in, width=2.5in]{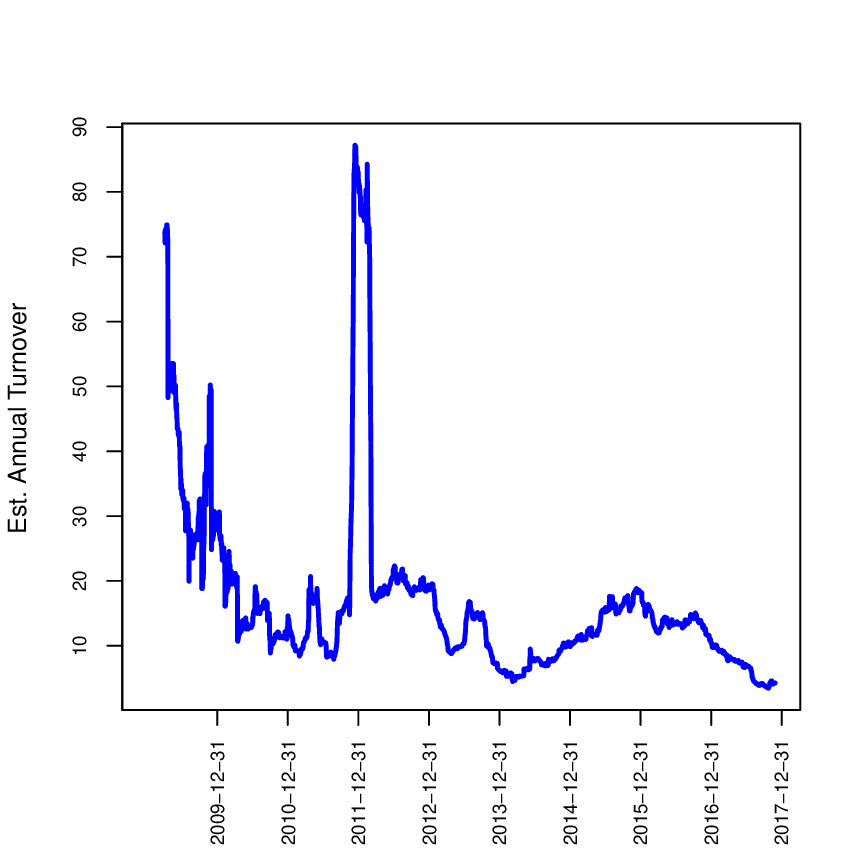}
 \end{tabular}
\caption{Annual turnover (average number of times spent per year) for on-chain spent bitcoins for 30 day and 90 day aggregated transaction data.}
\label{turnfig}
\end{figure}

In Figure \ref{dormfig}, we show the average dormancy from January 2009 until November 27, 2017 based on using daily bitcoin volume transaction and daily bitcoin days destroyed data under three scenarios of aggregation for the set of transactions in $\mathcal{S}$. These are when the transactions in $\mathcal{S}$ are aggregated over the periods of the previous one day, previous 30 days, or previous 90 days. The aggregation of trailing data does introduce a lag into the measured average dormancy but aggregation over a judicious time frame, we will focus on 30 days most, allows us to create an average metric that smooths out daily fluctuations and clarifies long term trends given bitcoin transaction value and bitcoin days destroyed are both cumulative measures. These are accompanied by Figure \ref{turnfig} which shows the inverse of the second two graphs multiplied by 365 to estimate the annual turnover for actively traded bitcoins.

Several trends are clear. First, the average dormancy, as well as the annual turnover, have both risen and fallen respectively over time since Bitcoin's inception. These trends were not monotonic and average dormancy especially seems to track Bitcoin price surges and falls. Since late 2012, average dormancy for bitcoins traded has rarely fallen below 20 days and only risen above 70 days in times of huge USD/BTC price volatility such as rapid appreciation of BTC relative to USD. In the current price surge, the 30 days aggregated average dormancy is around 100 days having peaked at almost 140 days earlier in August 2017. Likewise turnover is around 3.5 times per year and has only occasionally exceeded 20 times per year since late 2012.

The all-time peak, in early December 2011, is not attributable to one clear factor though it was the beginning of a price rally after a prior crash where transactions were at their second highest level while days destroyed plummeted. It also coincided with exposure in the mainstream media for Bitcoin during an episode of The Good Wife \cite{goodwife}, but the surge had started before the show aired. Subsequently,
 turnover experienced a nearly universally downward trend where annual turnover has now comfortably stayed between 10 to 20 times per year for nearly two years, until the recent Bitcoin appreciation to thousands of USD where it dropped to below 3 days briefly in August 2017.

\begin{figure}[ht]
\centering
 \begin{tabular}{cc}
 	 \includegraphics[height=2.5in, width=2.5in]{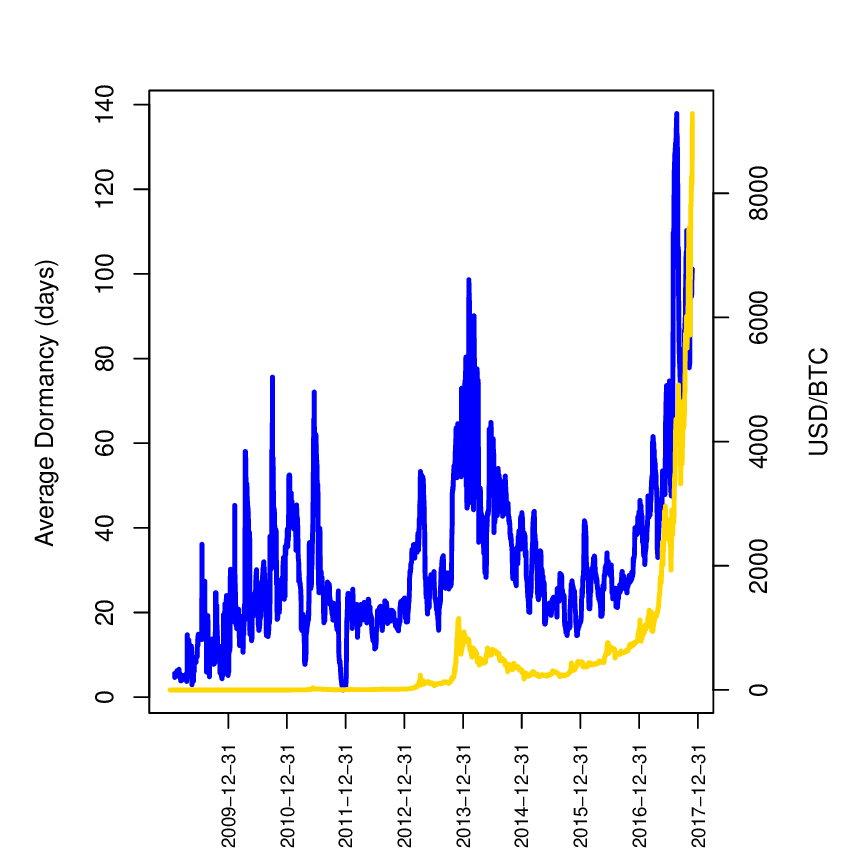}&
    \includegraphics[height=2.5in, width=2.5in]{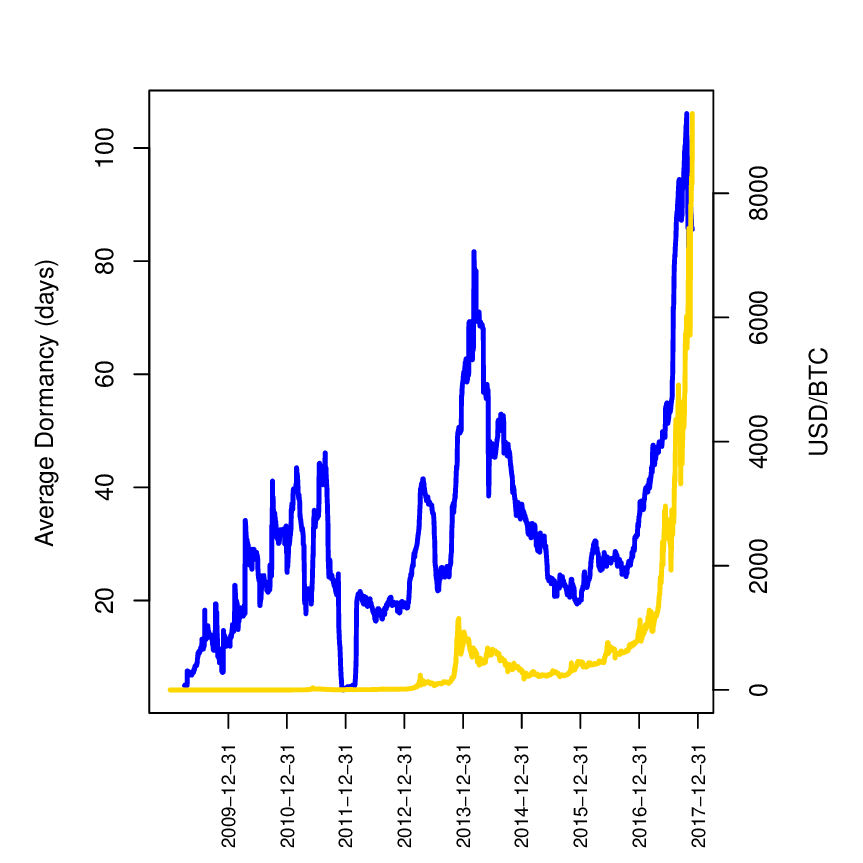}
 \end{tabular}
\caption{Average dormancy (30 day and 90 day aggregation; blue color) and the USD/BTC exchange rate (green color) tracked over time. Average dormancy days are tracked on the left axis, the USD/BTC exchange rate on the right axis.}
\label{exchangefig}
\end{figure}

\begin{figure}[ht]
\centering
 \begin{tabular}{cc}
 	 \includegraphics[height=2.5in, width=2.5in]{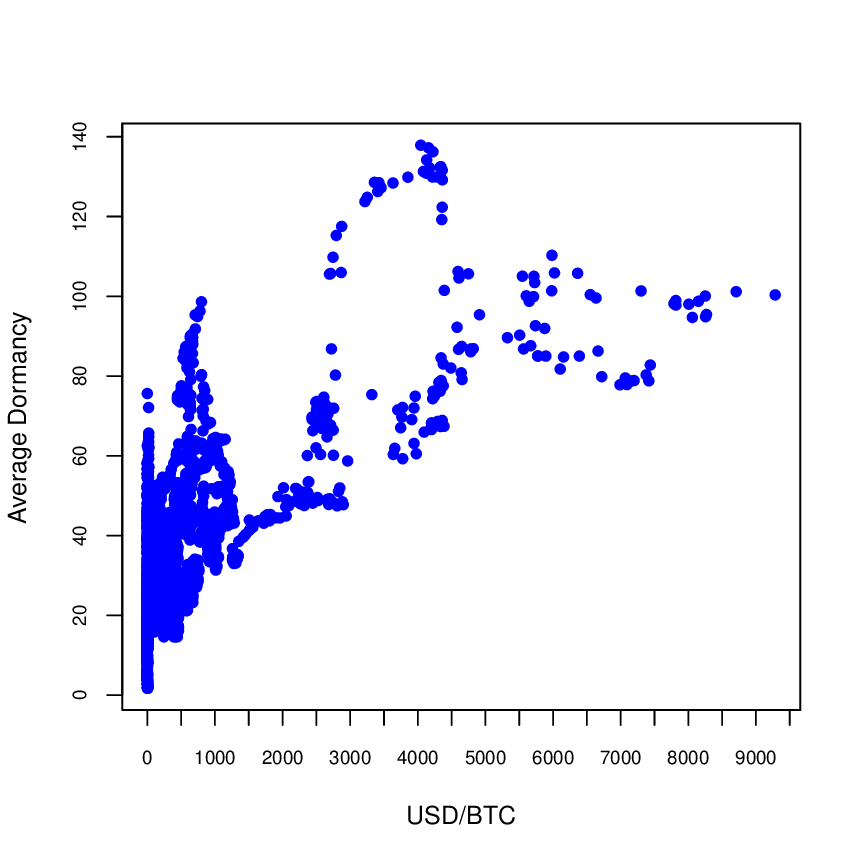}&
    \includegraphics[height=2.5in, width=2.5in]{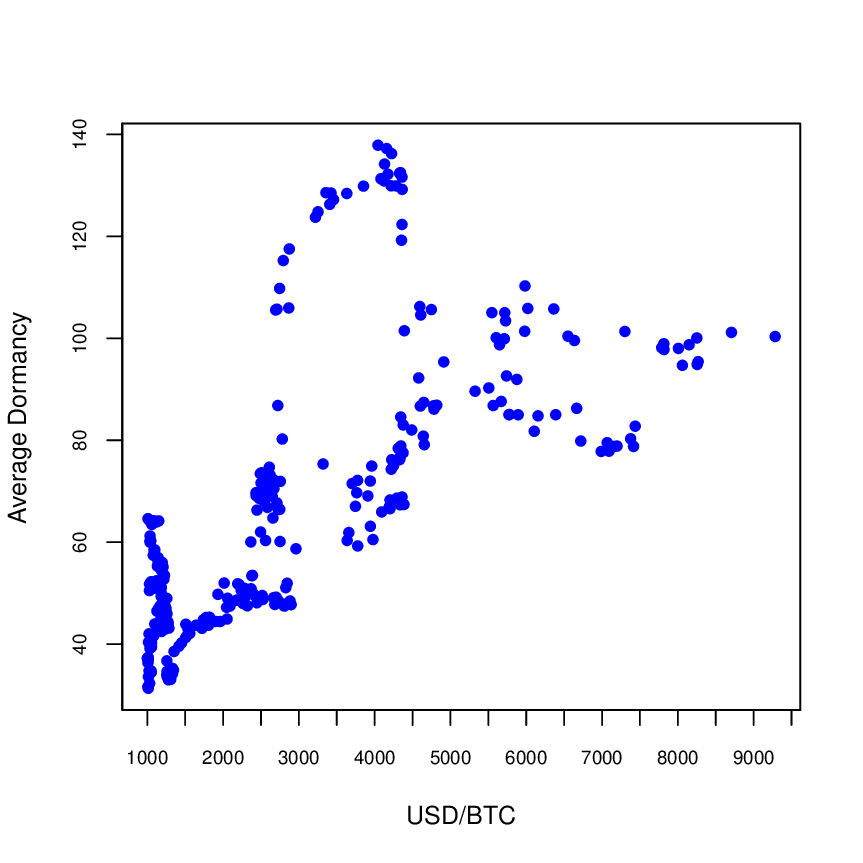}
 \end{tabular}
\caption{Average dormancy (30 day aggregation) vs. the USD/BTC exchange rate. The first graph represents the data over all time up until November 27, 2017 while the second graph represents the data only when USD/BTC exceeds \$1000.}
\label{dormUSD}
\end{figure}

Average dormancy's correlation to USD/BTC price trends is not a coincidence, though it is contextual. In Figure \ref{exchangefig} we see how closely the USD/BTC exchange rate and average dormancy track each other. In fact, given the shorter lag for the 30 day aggregation, we see up to the current price surge average dormancy and USD/BTC have tracked closely. This relation is much more similar to the USD/BTC exchange rate than either transaction volume or days destroyed demonstrate alone. The full context is shown, however, in the two graphs in Figure \ref{dormUSD}. The first shows average dormancy (30 day aggregated) vs. USD/BTC for all time while the second shows the same data but only when USD/BTC exceeds \$1000 / BTC. This clearly demonstrates the pattern is most apparent and consistent at times of relatively high value relative to the USD. Above the threshold of \$1000 / BTC there is a correlation of 0.68 between the two variables suggesting that at least 46\% of the variance in average dormancy is possibly directly related to the variance in the USD/BTC exchange rate at high values of exchange. 

There are at least two possibilities for the cause of this effect. One is that relatively valuable bitcoin heavily influences user spending behavior patterns. Another is a correlation induced between the exchange rate and average dormancy caused by a queuing effect on transactions related to the average block size. Around \$1500, the average size of blocks of transactions began to approach its maximum value of 1 MB. This meant that unlike most periods of Bitcoin's history, the latency for transaction confirmation and the average transaction fee began to climb starkly, possibly also causing the refrainment from unnecessary transactions by bitcoin holders. Both possible causes seem to have a correlation with average dormancy. The first surge in average dormancy to around 60-80 days in late 2013 and early 2014 was caused by the then unprecedented price rise that briefly exceed \$1000. Average block size then was about 0.2 MB and not a factor. More recently, however, the high price and long average dormancy, especially in excess of 80 days has only corresponded with average block sizes greater than about 0.8 MB.

The rise in average dormancy with price does not mean coins are being used less often but that alternate factors may affect trading in times of high prices. First, older coins that were hoarded are being traded, possibly due to the increased demand causing and following price rises that encourages those holding bitcoin to either sell their coins for fiat or exchange their newly valuable coins for goods and services. Second, the high prices and related volatility encourages trading and thus the increased parking of bitcoin on exchanges. While leaving bitcoin on an exchange is risky and not a best practice, actively trading bitcoin between accounts on an exchange is mostly off chain transactions. The bitcoin may not be destroyed until they are transferred from an address owned by the exchange to one owned by a trader when bitcoin are withdrawn. Thus an extended period of exchange trading makes many bitcoins look ``dormant'' as far as the Blockchain ledger is concerned.

This highlights an important conceptual understanding of average dormancy: it only measures the dormancy of coins used in active transactions. It does not tell us how dormant the overall monetary base of all bitcoins are. In fact, during lower price periods, a substantially smaller proportion of the money supply of bitcoin may be trading, but the average dormancy rate is lower since it only reflects the relatively short holding time of bitcoin users whose buying, selling, and trading activity is unimpeded by a relatively low cost per coin. Users who see bitcoin primarily as a store of value and want to take advantage of increasingly deflationary prices over time are more likely to sit out during these periods and trade less often, especially if their bitcoin were purchased at a higher USD (or other fiat currency) Bitcoin exchange rate.

This does seem to confirm, however, that users who have held coins for long periods without trading are more likely to trade these coins as the value of Bitcoin versus the USD rises. This may also reflect the primacy of exchanges as dominant venues of bitcoin trading and transactions. Thus average dormancy is increased by the `emergence' of coins that are being traded in various off-chain transactions such as being stored and traded on exchanges, coins being removed from exchange cold storage to enable withdrawals or other trades, or those that had been obtained via private transactions such as surrendering a private key. On balance, it seems a high price for Bitcoin relative to fiat seems to encourage less hoarding, not more.

\section{Days destroyed analyzed by Little's Law}
A short discourse on Little's Law, despite being largely inapplicable to long-term bitcoin trading data, adds an additional detail to days destroyed as well. Little's Law is a commonly used and basic equation in queuing theory. Though it has existed in different forms throughout time in different fields being applied to more narrow problems, its most general and widely used form was first proposed, without proof, by Philip Morse \cite{little1} in 1958 and given a rigorous proof by John Little \cite{little2} in 1961. Little's Law relates the average number of items in a queue, $L$ \cite{littleexample} (bank queue line customers, cars at a toll booth, even Drake's equation for the estimated number of intelligent species \cite{seti}) to the average arrival rate at the queue, $\lambda$, and the average wait time in the queue $W$.

\begin{equation}
L = \lambda W
\label{littleslaw}
\end{equation}

It is a simple matter to compare equations \ref{ddeq3} and \ref{littleslaw} to see that the relationship between days destroyed and the Bitcoin transaction volume is an expression of Little's law where the average wait time, $W$, is replaced by $\langle t \rangle$, the arrival rate, $\lambda$, is replaced by $B$, the transaction volume, and the number of items in the queue, $L$, is replaced by days destroyed, $D$.

To understand this perspective, one must abstract the basic spend activity of on-chain transactions as trades between two addresses where the recipient, on average, does not trade the newly created bitcoin for $\langle t \rangle$ days. In this conceptualization, the days destroyed has an alternate interpretation: it is the average size of the total pool of bitcoins spent in on-chain transactions during a time horizon equal to the average dormancy.

This interpretation, however, is flawed and not realistic for long-term bitcoin trading patterns. This is primarily because Little's Law requires the underlying random processes, in particular the bitcoin transaction volume and the dormancy rate, to be stationary: having the same distribution and moments (mean, standard deviation, etc.) over time.  Financial market data is almost never stationary, even under short time periods. It would require a relatively level trading volume and constant dormancy rate for Little's Law to work for the alternate interpretation of days destroyed to be fully valid. Therefore measurements of the pool of bitcoins used in on-chain transactions by virtue of days destroyed may be approximate only for limited time series, such as the period of average dormancy under minimal exchange rate price volatility, but definitely not for months or years.

\section{Non-normality of bitcoin days destroyed}

\begin{figure}[ht]
\centering
\includegraphics[width=3in]{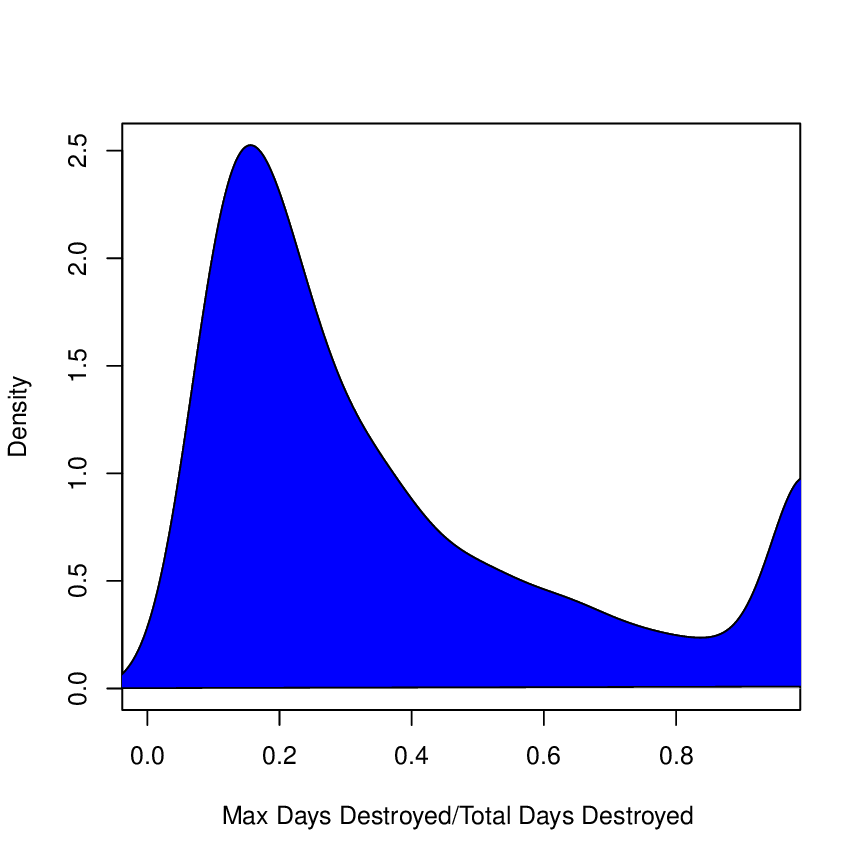}
\caption{Probability density plot of the proportion of the daily days destroyed that is accounted for by the largest transaction from the Genesis block until November 27, 2017.}
\label{DDskew}
\end{figure}

Previously, we have discussed bitcoin days destroyed and trading volume in terms of aggregate sums over time related by the average dormancy. While this analysis is mathematically accurate, using the mean or sum of any of the variables requires caution due to the nature of the data. Like all financial markets \cite{mandelbrot,market2,market3,hft}, the transactions in Bitcoin, both in volume and days destroyed, show a heavily skewed character where large transactions can be of an almost arbitrarily large size, implying a huge variance. This is an issue not unique to Bitcoin, but a well known issue dealing with financial market data.  In fact, the skewed nature of the distribution of days destroyed is shown in Figure \ref{DDskew} where the proportion of daily bitcoin days destroyed accounted for by the largest transaction value in days destroyed can be as high as almost 100\% while rarely dipping below 10\%. The median proportion of total daily days destroyed accounted for by the largest transaction for 2017 through November is 17\%--almost one-fifth of days destroyed on average is accounted for by the largest transaction measured by days destroyed.

The actual distribution of days destroyed or transaction volume is outside the scope of this paper but is possibly a power law, skewed exponential, or another such long tail which can be verified by the appropriate statistical tests. While this does not invalidate the analysis it shows that averages, such as dormancy rate, can be affected by large transactions in days destroyed or bitcoin transaction volume that are not accompanied by large increases in the other variable. This can be large transaction volumes of coins that have moved recently or huge days destroyed transactions based on relatively small transfers of aged coins.

For example, if the one million or so BTC attributed to addresses purportedly owned by the Bitcoin founder(s) Satoshi Nakomoto ever moved, there would be a massive spike in days destroyed—about eight times larger than the previous largest days destroyed transaction and twice as large as the total days destroyed for the highest day ever. Based on 2017 median 30 day aggregate transaction and days destroyed volumes, the measured average dormancy would soar past a year,  skewing these variables way out of proportion to the normal network activity.

Aggregation, while smoothing out daily fluctuations, does not remove these effects. Removing the largest values, while it may provide some balance, also cannot totally remove the skew. If the size distributions of days destroyed or transaction volume hold a power law character, the nature and proportion of skewed transactions will always exist no matter which proportion of the top values are removed.

\section{Conclusion}

The average dormancy of actively traded bitcoin, easily generalizable to almost any cryptocurrency, is a valuable new variable that ties together previous concepts of transaction volume into a coherent picture. The understanding it brings of the relative dormant period of bitcoins before use allows us to understand the aggregate behavior of the bitcoin user base in a more intuitive fashion than bitcoin days destroyed and a less noisy fashion than transaction value. While discussed less in this paper, the inverse of the average dormancy, the turnover, can provide a pseudo-velocity metric that shows how often bitcoins circulate through its on-chain economy over a given time period. Granted, this is probably an underestimate of the true usage of bitcoin since it only measures on-chain transactions but it is interesting to note that both it and average dormancy point to the likelihood that the more valuable Bitcoin is relative to fiat, the \emph{lower} the overall turnover is on the Blockchain due to a variety of possible reasons. If Bitcoin is only seen as a store of value, maybe this is not an issue but if it wants to be used and considered as any other currency, an appreciable turnover of on-chain transactions seems preferable. The Blockchain as an innovation is what in part distinguishes Bitcoin from past, failed digital currencies.

While average dormancy and turnover may help us understand how bitcoin transactions relate to bitcoin's use and circulation, neither measure should be mistaken for an exact analogue of monetary velocity. Neither dormancy or turnover play such a role mediating a relationship with the relative value of Bitcoin or the overall base of mined bitcoin, despite what they show about bitcoins in active use in transactions. Monetary velocity, while often defined as the number of transactions per currency unit per unit time, is a relation that connects the money supply, economic activity, and the price inflation. Also, to follow a relation such as the exchange equation, Bitcoin would have to have a primarily utilitarian, not store of value, interpretation similar to the Quantity Theory of Money.

The unique nature of Bitcoin and its current and future uses as a currency or store of value make it an ideal candidate for studying it from a variety of perspectives. In order to better understand Bitcoin and its uses over time, a variety of new measures such as bitcoin days destroyed have been created to describe activity on the Blockchain. The two new measures proposed in this paper, average dormancy and turnover of actively traded coins, assist in these analyses by giving aggregate and easily derivable properties of the bitcoin in use. It is hoped that these measures will become better understood and researched in order to enhance our understanding of this wonderful innovation and its future evolution.


\ledgernotes

\section{Acknowledgement}

The author would like to thank OXT for providing the raw data sets for days destroyed (total and maximum). The author would also like to thank the anonymous peer reviewers for their helpful insight and comments.

\section{Author Contributions}

The sole author of this paper was its only contributor.

\section{Conflict of Interest}

None



\thispagestyle{pagelast}

\end{document}